\documentclass{PoS}
\usepackage{amsmath}

\newcommand{\beq}{\begin{eqnarray}}
\newcommand{\eeq}{\end{eqnarray}}

\title{T-odd Transverse Momentum Distributions in Quark Models}

\ShortTitle{T-odd TMDs in Quark Models}

\author{\speaker{A.~Courtoy}\\
        INFN-Sezione di Pavia\\
27100 Pavia, Italy.\\
        E-mail: \email{aurore.courtoy@pv.infn.it}}

\abstract{
We present a general formalism for the evaluation of time-reversal odd parton distributions. This formalism is  applied to the evaluation of the two T-odd TMDs allowed at the twist-2 level, i.e., the Sivers and the Boer-Mulders functions. We have performed the calculation for two different models of proton structure: a non relativistic constituent quark model and the MIT bag model. The results obtained in both models are compared. So are the two T-odd functions. We comment on the fulfilment of the Burkardt sum rule as well as on the current status of T-odd TMDs in phenomenology.
}

\FullConference{Light Cone 2010 - LC2010\\
		June 14-18, 2010\\
		Valencia, Spain}
\begin{document}

Our knowledge on the hadron structure is  incomplete. We, for example, know that the parton distribution functions describe the 1-D structure of hadrons. At leading order, the pdfs are three: number density, helicity and transversity. However the experimental knowledge on the latter is rather poor as it is a chiral-odd quantity not accessible through fully inclusive processes.
Therefore, Semi-Inclusive DIS experiments, where a final hadron is detected, have been proposed to extend our knowledge on this function. This extension from fully inclusive to semi-inclusive processes implies a generalization of the distribution functions. That is,  if one wants to study the transverse momentum distribution of the produced hadron, one has to account for transverse motion of quarks. In fact,  we now know  that non-perturbative effects of the intrinsic transverse momentum $\vec{k}_{T}$ of the quarks inside the nucleon may induce significant hadron azimuthal asymmetries \cite{Mulders:1995dh, Cahn:1978se}.

It is in this context that the Sivers and the Boer-Mulders functions were defined. 
Transverse Momentum Dependent pdfs (TMDs) are the set of functions that depend on both the Bjorken variable and the intrinsic transverse momentum of the quark.\footnote{They also depend on the scale $Q^2$, like the pdfs.} 
Their number is fixed by the number of scalar structures allowed by hermiticity, parity and time-reversal invariance. 
 However, the existence of final state interactions allows for time-reversal odd functions~\cite{Brodsky:2002cx}. In effect, by relaxing this constraint, one defines two additional functions, namely, the Sivers and the Boer-Mulders function. These functions are related, respectively, to single spin and azimuthal asymmetries, and are therefore important in our quest for the understanding of the proton spin.
\\

The Sivers function, $f_{1T}^{\perp {q} } (x, k_T)$ \cite{sivers}, and the Boer-Mulders function,  $h_{1}^{\perp {q} } (x, k_T)$ \cite{Boer:1997nt},  are formally defined,  according to the Trento convention \cite{trento}, for the quark of flavor ${q}$, through the following expression\footnote{$a^\pm = (a_0 \pm a_3)/ \sqrt{2}$.}:
\begin{eqnarray}
f_{1T}^{\perp {q}} (x, {k_T} ) 
& = &- {M \over 2k_x} \,\int \frac{d\xi^- d^2\vec{\xi}_T}{(2\pi)^3} \ e^{-i(xp^+ \xi^- -\vec{k}_T\cdot \vec{\xi}_T)} \,\nonumber\\
& \times &
{1 \over 2} \sum_{S_y=-1,1} \, S_y \, \langle P S_y\vert \overline \psi_{q} (\xi^-, \vec{\xi}_T)\, {\cal L}^{\dagger}_{\vec{\xi}_T}(\infty, \xi^-)\, 
\gamma^+\,{\cal L}_0(\infty,0) \psi_{q}(0,0) \vert P S_y\rangle \, + \mbox{h.c.} \, ,
\label{siv-def-op}
\eeq
taking the proton polarized  along the $y$ axis; and
\beq
h_{1}^{\perp {q}} (x, k_T) 
& = & - {M \over 2 k_x} \,\int \frac{d\xi^- d^2\vec{\xi}_T}{(2\pi)^3} \ e^{-i(xp^+ \xi^- -\vec{k}_T\cdot \vec{\xi}_T)} \,\nonumber\\
& \times &
{1 \over 2} \sum_{S_z=-1,1} \langle P S_z\vert \overline \psi_{q} (\xi^-, \vec{\xi}_T)\,  {\cal L}^{\dagger}_{\vec{\xi}_T}(\infty, \xi^-)\, 
\gamma^+\gamma^2\gamma_5\,{\cal L}_0(\infty,0) \psi_{q}(0,0) \vert P S_z\rangle \, + \mbox{h.c.} \,,
\label{bm-def-op}
\eeq
where $ {\vec S} $ is the spin of the target hadron. The normalization of the covariant spin vector is $S^2 = -1$,
$M$ is the target mass, $\psi_{q}(\xi)$ is the quark field
and  ${\cal L}_{\vec{\xi}_T}$ is the gauge link.\footnote{The gauge link is defined as
$
 {\cal L}_{\vec{\xi}_T}(\infty, \xi^-)= {\cal P} \mbox{exp}\left( -ig\, \int_{\xi^-}^{\infty} \, A^+(\eta^-,\vec{\xi}_T)\, d\eta^-\right)\,, 
$
where $g$ is the strong coupling constant. This definition holds in covariant (non singular) gauges, and in SIDIS processes, as the definition of T-odd  TMDs is process dependent.} The gauge link contains a scaling contribution which makes the T-odd TMDs non vanishing in the Bjorken limit.

The difference between the two functions is clearly  physically transparent from  Eqs. (\ref{siv-def-op}) and (\ref{bm-def-op}). 
The BM function counts the transversely  polarized quarks, hence the Dirac operator $\gamma^+\gamma^2\gamma_5$ in Eq. (\ref{bm-def-op})) in an unpolarized proton. On the other hand, the Sivers function counts the unpolarized quarks, hence the Dirac  operator $\gamma^+$ in Eq. (\ref{siv-def-op}), in a transversly polarized proton, i.e. the explicit transverse component $S_y$ in Eq. (\ref{siv-def-op}). If there were no  scaling contribution of the gauge link, the two T-odd functions $f_{1T}^{\perp {q} } (x, k_T)$ and $h_{1}^{\perp {q} } (x, k_T)$ would be identically zero.

\section{The Interaction: MIT bag vs. NRCQM}

\begin{figure}
\begin{center}
\includegraphics[height=3.5cm]{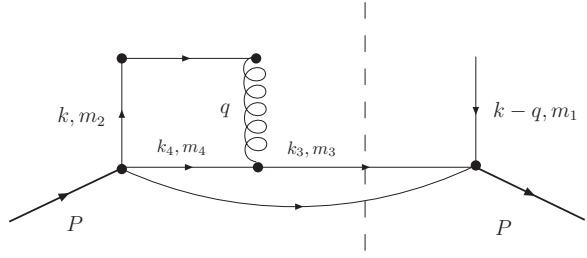}%
\caption{
Contributions to the T-odd TMDs.
}
\label{feyn}%
\end{center}
\end{figure}
The general formalism for time-reversal odd functions in quark models has been presented in Ref.~\cite{Courtoy:2008vi}. This formalism implies the development into free-quark states and requires that we work within  the Impulse Approximation. None of the quark model we have used contain gluonic degrees of freedom, therefore the Final State Interactions are consistently introduced through a One Gluon Exchange, as depicted in Fig.~\ref{feyn}. The  expression for the Sivers and the BM functions are similar in any quark model: the Dirac structure dictates the spin structure of the quark operator defining the interaction $V_{m_1m_2m_3m_4}$ ; 
 then  the resulting spin-flavor-color matrix elements dictate the spin combinations allowed by such a structure and reflects also  our modelling  of the proton. For instance, in the MIT bag model~\cite{Jaffe:1974nj}, one gets,\footnote{The expressions as well as the subscripts in a $[\, ]$ refer to the BM function.}
\beq
f_{1T} [h_{1}]^{\perp {q}} (x, k_T)&=&-2\Im[\,]\,  i g^2\,\frac{M\, E_P}{k_{x}}\, 
\int \frac{d^2 \vec{q}_T}{(2\pi)^5}
\int \frac{d^3 k_3}{(2\pi)^3}
\sum_{m_1,m_2,m_3, m_4} 
C_{{q} f[h]}^{m_1m_2, m_3m_4}\, V(\vec{k}, \vec{k}_3, \vec{q}_T)_{m_1m_2, m_3m_4}
 \quad,
 \nonumber\\
\label{bm-bag-ready}
\eeq
where no further recoil of the target is considered. 
The imaginary part is taken only for the Sivers function ; it is a consequence of re-expressing  the transverse proton spin component in an helicity basis. 
 The spin combinations at the quark level appear clearly from the expressions
  \beq
  C_{{q} f[h]}^{m_1m_2, m_3m_4} &=&{1 \over 2} \sum_{S}\,C_{{q}S}^{m_1m_2, m_3m_4}\nonumber\\
  &=&\sum_{\beta} \,T^a_{ij}T^a_{kl}\,\langle P S_z=1|  b_{{q} m_1}^{i\dagger}b_{{q} m_2}^{j}b_{\beta m_3}^{k\dagger}b_{\beta m_4}^{l} |P  S_z=-1\rangle\nonumber\\
&&\left[{1 \over 2} \sum_{S_z=1,-1}\,\sum_{\beta} \,T^a_{ij}T^a_{kl}\,\langle P S_z|  b_{{q} m_1}^{i\dagger}b_{{q} m_2}^{j}b_{\beta m_3}^{k\dagger}b_{\beta m_4}^{l} |P  S_z\rangle\right]
\quad,
\label{me}  
\eeq
whose calculation is performed, here, assuming $SU(6)$ symmetry,  for ${q}=u,d$. The interaction $V(\vec{k}, \vec{q}_T)$ is here evaluated using
the properly normalized fields for the quark in the bag \cite{Jaffe:1974nj}  given in terms of the quark wave function in momentum
space, which reads
\begin{eqnarray}
\varphi_m(\vec{k})&=&i\, \sqrt{4\pi}\, N\, R_0^3 
\begin{pmatrix}
 t_0(|\vec{k}|) \chi_m
\\
{\vec{\sigma} \cdot \hat{k}}\,t_1(|\vec{k}|)\, \chi_m
\end{pmatrix} \quad,
\label{bagwf}
\end{eqnarray}
with the normalization factor $N$. The interaction is then
\beq
V(\vec{k}, \vec{k}_3, \vec{q}_T)_{m_1m_2, m_3m_4} &=&
\frac{1}{q^2}\, \varphi^{\dagger}_{m_1} (\vec{k}-\vec{q}_T)\,\gamma^0 \gamma^+
\Gamma^{f[h]}\,
 \varphi_{m_2} (\vec{k})
\varphi^{\dagger}_{m_3} ( \vec{k}_{3})\,\gamma^0 \gamma^+\, \varphi_{m_4} 
(\vec{k}_{3}-\vec{q}_T)\,,
\label{int-bag}
\eeq
where $\Gamma^{f[h]}=1$ or $\gamma^2\gamma_5$ for, respectively, the  Sivers and the Boer-Mulders  function.
The expressions in the NRCQM are similar to the one in the bag model. They read, with $\Psi$ the intrinsic proton wave function,
\beq
f_{1T} [h_{1}]^{\perp {q}} (x, k_T)&=&-2 i g^2\,\frac{M\, E_P}{k_{x}}\, 
\int \frac{d^2 \vec{q}_T}{(2\pi)^5}
\,
\sum_{m_1,m_2,m_3, m_4}\nonumber\\
&&\int \,d\vec{k}_1 \, d\vec{k}_3\,(2\pi)^3 \delta(k_1^+-xP^+) \delta(\vec{k}_{1\perp}+\vec{q}_{\perp}-\vec{k}_{\perp})\nonumber\\
&&{1 \over 2} \sum_{S}\,\sum_{\beta}\sum_{ijkl}
\delta_{\mbox{\tiny QN}} \,\,
\Psi^{\dagger}_{r \, S}
\left ( \vec k_1, \{m_1,i,{q}\}; \, \vec k_3, \{m_3,k,\beta \};
\, - \vec k_3 - \vec k_1,  m_n  \right )\,\,T^a_{ij}  T^a_{kl}
\nonumber \\
& &
V(\vec k_1, \vec k_3, \vec q)_{m_1m_2, m_3m_4}
\Psi_{r \, S}
\left (\vec k_1 + \vec q, \{m_2,j,{q}\}; \, \vec k_3 -
\vec q, \{m_4,l,\beta \};
\, - \vec k_3 - \vec k_1,  m_n  \right )
 \, .
 \nonumber\\
\label{bm-cqm-ready}
\eeq
The intrinsic proton wave function explicitly depends on the momenta so that there is no possible factorization of the spin-flavor-color matrix elements, like in Eq.~(\ref{bm-bag-ready}).
The interaction here reads
\beq
V(\vec k_1, \vec k_3, \vec q)_{m_1m_2m_3 m_4}
&=& \frac{1}{q^2}\, \bar u_{m_1}(\vec{k}_1)
\,\gamma^+\Gamma^{f[h]}\,
u_{m_2}(\vec{k}_1 + \vec q )
\,  \bar u_{m_3}(\vec{k}_3)
\,\gamma^+\,
u_{m_4}(\vec{k}_3-\vec{q})
\quad,
\label{int}
\eeq
with $u(\vec{k})$ the four-spinor of the free quark states.

\section{Results with an $SU(6)$ proton wave function}

In Refs.~\cite{Courtoy:2008vi, Courtoy:2008dn, Courtoy:2009pc}, we have evaluated both T-odd TMDs in both the MIT bag model and a non relativistic Constituent Quark Model (NRCQM) using $SU(6)$ symmetry for the proton. 

In the former case,  the qualitative results are  directly reflected  from the calculation of the coefficients (\ref{me}).
In effect, through those coefficients it is possible to reconstruct what happens at the level of the quark helicity in a perfectly transparent way. 
In the latter case,  one first has to re-express Eq.~(\ref{bm-cqm-ready}) in terms of the formal expression of  the proton state. In spectroscopic notation and with the Jacobi coordinates, one has
\beq
 | ^2 S_{1/2} \rangle_S&=& \frac{
 e^{-\left(k^2_{\rho} +k_{\lambda}^2\right)/\alpha^2}
 }{\pi^{3/2} \alpha^3} | \chi \rangle_S~,
\label{wf}
\eeq
where $  | \chi \rangle_S $ is the standard $SU(6)$ vector describing the spin-flavor structure of the proton.

Therefore, in the MIT bag model, the r\^ole of each contribution can be followed and evaluated. Similar conclusions can be driven from the NRCQM calculation but the expression are slightly more intricate due to the momentum dependence of the proton wave function. It is however worth noticing that the main difference between the two model's philosophy is that the interference term giving rise to a non-zero Sivers function arises either due the MIT bag wave function or to the 4-spinors of the free quark states, in the CQM approach. This appears clearly in comparing the expressions Eqs.~(\ref{int-bag}, \ref{int}) for the interaction.
There are only 2 spin combinations contributing to the Sivers function.
The dominant contributions to this function comes from the spin-flipping of the quark interacting with the photon, i.e. the  $Y$ term in the MIT bag calculation
\begin{eqnarray}
f_{1T}^{\perp {q}} (x, k_T)&=&\frac{g^2}{2}\frac{ME_P}{k^x} \,C^2
\int
\frac{d^2q_\perp}{(2\pi)^2}\frac{1}{q^2} [ C_{q}^{-+} Y(\vec
q_\perp,k_T) + C_{q}^{+-} U(\vec q_\perp,k_T)]~,\label{sivf}
\end{eqnarray}
with $C$ a normalization factor, $C_{q}$ a weighting spin-flavor-color factor resulting from the matrix elements~(\ref{me}) and where $Y/U(\vec
q_\perp,k_T)$ include the momentum dependent part.\footnote{See Eqs.~(8-9) of Ref.~\cite{Courtoy:2008dn}.}

On the other hand, there are more spin combinations for the BM function.\footnote{See Eq.~(13) of Ref.~\cite{Courtoy:2009pc}.
} The first reason is that both non-flipping and double-flipping terms are important. The second reason is the sum over the two spin states, i.e. $S_z=-1,1$.
Due to the spin-flavor-color coefficients, i.e., due to the
$SU(6)$ symmetry assumption, the non-flipping term {\it weights} more than the double-flipping contribution.
In effect, the latters are governed by the product of the
two lower components of the bag wave function  which encodes the most relativistic contribution
arising in the MIT bag model.
They turn out to be a few orders of magnitude smaller
than the dominant ones, arising from the interference between
the upper and lower parts of the bag wave function.
This  also happens if a proper non relativistic reduction of the gauge link, suitable for CQM calculations, is performed, justifying then the non relativistic approximation.

On Fig.~\ref{cqm-bag-both} we show  both the Sivers and the Boer-Mulders functions for $u$ and $d$ quarks in both the CQM and MIT bag model, with the value $\alpha_s(\mu_0^2)/(4 \pi) \simeq 0.13$~\cite{Traini:1997jz}. We next explain why we choose such a value for the strong coupling constant.

\section{From Model Calculation to QCD}

\begin{figure}
\begin{center}
\includegraphics[height=6.6997cm,width=7.9737cm]{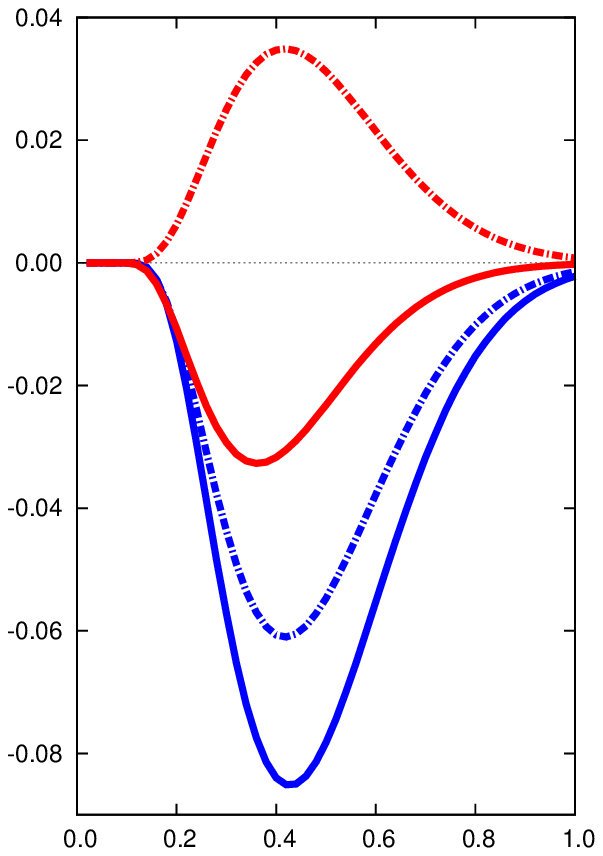}%
\includegraphics[height=6.6997cm,width=7.9737cm]{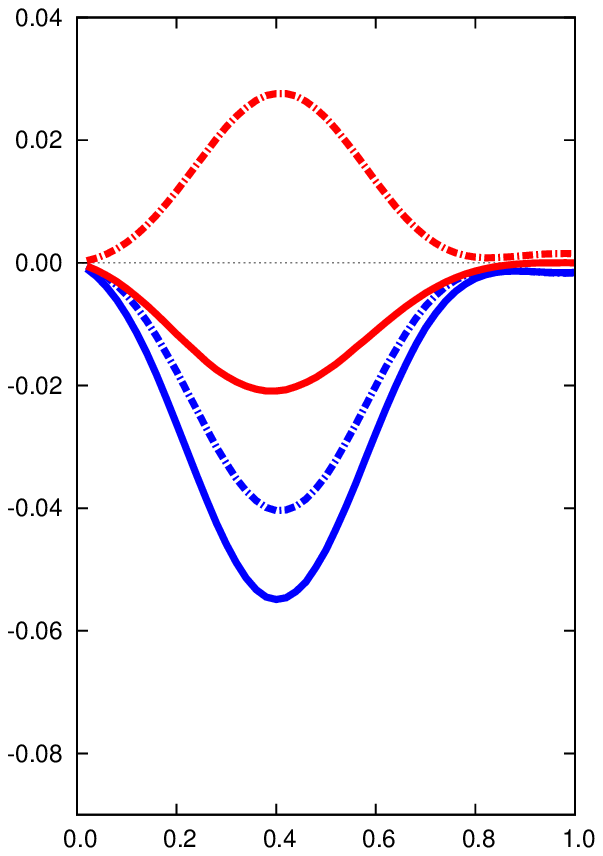}%
\caption{
Comparison of both the Sivers (red) and the Boer-Mulders (blue) functions in the NR Constituent Quark Model (left) and the MIT bag model (right). Small dashed curves represent the $d$-quark distributions ; full curves the $u$-quark. 
}
\label{cqm-bag-both}%
\end{center}
\end{figure}

Model calculations are important for their inputs to our current knowledge about TMDs. Most of the models present in the literature are relevant at scales as low as $0.1$ GeV$^2$. This scale is identified by matching the experimental value of the second moment of the unpolarized pdf with the result obtained within the model. It moreover represents the energy from which we assume perturbative QCD to be applicable. 

 Nevertheless, the latter argument is controversial but, so far~\cite{Cherednikov:2010uy}, that's the best we can do. The problem is a bit more tricky when one goes to transverse momentum dependent distributions, like we are doing here, as the TMD community hasn't come through the TMD QCD evolution's complexity  yet. Basically, by  trying to link model calculation with QCD, one faces the 2 problems: a  "too" low initial scale and no workable evolution equations. 

In Refs.~\cite{Courtoy:2008vi, Courtoy:2008dn} we have nevertheless tried our hand by comparing our results for the first moment of the Sivers function with the extraction from the HERMES data by evolving, up to the experimental scale, our low-scale result. The trend followed by both curves is similar but it is clear that both results can be improved; e.g. the model calculation can be improved by evolving properly ; the parametrization also, by, e.g., including more data. 

In a more naive spirit, we here give a comparison of the Boer-Mulders function calculated in both the MIT bag model and a NRCQM with a very first extraction from the unpolarized SIDIS data  on the $\cos 2 \phi$ asymmetry from  COMPASS and HERMES~\cite{Barone:2009hw}. The latter  is plotted in green in Fig.~\ref{bm-fig} with no error bands as the authors of Ref.~\cite{Barone:2009hw} have estimated the theoretical errors to be bigger than the errors from the data. This justifies that they would consider neither the theoretical nor the experimental ones. 
A proper evolution of the model results to the experimental scale would push the distributions towards small $x$, as happens with pdfs. Also, we expect the magnitude of the distribution to slightly decrease when evolved to higher scales. However, for the two reasons we have given above, no conclusion can be driven concerning the behavior in $Q^2$ of the TMDs.
%
\begin{figure}[tb]
\begin{center}
\includegraphics[height=6.6997cm,width=7.9737cm]{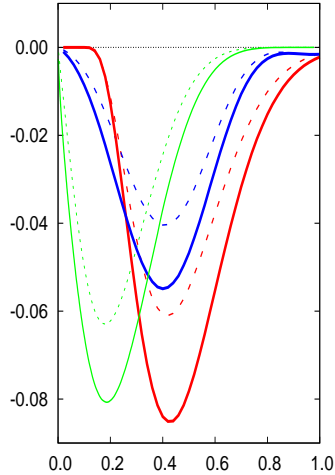}%
\caption{
The Boer-Mulders function in both the  NR Constituent Quark Model (red) and the MIT bag model (blue). Small dashed curves represent the $d$-quark distributions ; full curves the $u$-quark.  The green curves corresponds to the parameterization at SIDIS'scales (see text).
}
\label{bm-fig}%
\end{center}
\end{figure}

\section{The Burkardt Sum Rule}

There exist many model calculations as well as many variations on some models ; we refer to Refs.~\cite{Courtoy:2008vi, Courtoy:2008dn, Courtoy:2009pc} for bibliography. We nevertheless think it is worth comparing our results with a recent evaluation of the T-odd TMDs in a light-cone quark model~\cite{Pasquini:2010af}. As it could be expected, the results obtained by overlap of Light Cone Wave Functions are consistent with ours, e.g., predictions similar in shape and non proportionality between the $u$ and the $d$ quark distributions. The latter point is of great interest as it is related to the Burkardt Sum Rule~\cite{Burkardt:2004ur}. 

Apart from the scale input, model calculations are important in that some first principle properties can either be checked or lead to constraints. By first principle properties we refer, in the distribution functions framework, to sum rules. The Burkardt sum rule is related to momentum conservation: as there is no net transverse momentum in the proton, the sum over all partons of the first moment of the Sivers function should vanish~\cite{Burkardt:2004ur}.
In Refs.~\cite{Courtoy:2008vi, Courtoy:2008dn} we have shown that the Burkardt sum rule is fulfilled at a level of a few percent for both the MIT bag model and the CQM. This "discrepancy" cannot be fixed by correcting the support and it has therefore to be understood as a shortcoming of the two models. Moreover, this shortcoming should somehow be related to momentum conservation.
%
%
As the bag states are not good momentum eigenstates, we do not expect the bag model results to fully satisfy the Burkardt Sum Rule. 
Also, the approximations that we have used in the CQM calculations result in a breaking of the momentum conservation. Namely, the Impulse Approximation as well as the non relativistic reduction forbids momentum to be  fully conserved. This defect could not possibly be fixed by relaxing the non relativistic reduction or by applying the techniques of Ref.~\cite{Traini:1997jz}. It is actually the Impulse Approximation that renders this recovery impossible in our approach.

On the other hand, the LCWF fully satisfies the Burkardt  sum rule. While it is fully relativistic, it also makes use of the Impulse Approximation, but, in this case, it allows for momentum conservation. 
The difference between  our approach~\cite{Courtoy:2008vi} and the approach of Ref.~\cite{Pasquini:2010af} lies in the dynamical form, e.g. we work in the instant form while the LCWF approach is defined in the front form. Focusing on this difference, we conclude that the results of using an approximation depends on the dynamical form.



\acknowledgments

The author is grateful to V. Vento and S. Scopetta for former collaboration during my PhD at the Universidad de Valencia. I would like to thank B. Pasquini for fruitful discussions and V. Vento for his advices.


\begin{thebibliography}{99}



\bibitem{Mulders:1995dh}
  P.~J.~Mulders and R.~D.~Tangerman,
  Nucl.\ Phys.\  B {\bf 461} (1996) 197
  [Erratum-ibid.\  B {\bf 484} (1997) 538]
  [arXiv:hep-ph/9510301].
  
\bibitem{Cahn:1978se}
  R.~N.~Cahn,
  Phys.\ Lett.\  B {\bf 78} (1978) 269.
  
\bibitem{Brodsky:2002cx}
  S.~J.~Brodsky, D.~S.~Hwang and I.~Schmidt,
  Phys.\ Lett.\  B {\bf 530} (2002) 99


  \bibitem{sivers}
  D.~W.~Sivers,
  Phys.\ Rev.\  D {\bf 41}, 83 (1990),
  Phys.\ Rev.\  D {\bf 43}, 261 (1991).
  
\bibitem{Boer:1997nt}
  D.~Boer and P.~J.~Mulders,
  Phys.\ Rev.\  D {\bf 57} (1998) 5780
  [arXiv:hep-ph/9711485].
  
  \bibitem{trento}
  A.~Bacchetta, U.~D'Alesio, M.~Diehl and C.~A.~Miller,
  Phys.\ Rev.\  D {\bf 70}, 117504 (2004)
  [arXiv:hep-ph/0410050].
  
  
  

\bibitem{Jaffe:1974nj}
  R.~L.~Jaffe,
  Phys.\ Rev.\  D {\bf 11} (1975) 1953.
  


\bibitem{Courtoy:2008vi}
  A.~Courtoy, F.~Fratini, S.~Scopetta and V.~Vento,
  Phys.\ Rev.\  D {\bf 78} (2008) 034002
  [arXiv:0801.4347 [hep-ph]].


\bibitem{Courtoy:2008dn}
  A.~Courtoy, S.~Scopetta and V.~Vento,
  Phys.\ Rev.\  D {\bf 79} (2009) 074001
  [arXiv:0811.1191 [hep-ph]].

\bibitem{Courtoy:2009pc}
  A.~Courtoy, S.~Scopetta and V.~Vento,
   Phys.\ Rev.\  D {\bf 80} (2009) 074032
  [arXiv:0909.1404 [hep-ph]].

  
\bibitem{yuan}
  F.~Yuan,
  Phys.\ Lett.\  B {\bf 575} (2003) 45
  [arXiv:hep-ph/0308157].

  
\bibitem{Traini:1997jz}
  M.~Traini, A.~Mair, A.~Zambarda and V.~Vento,
  Nucl.\ Phys.\  A {\bf 614} (1997) 472.
  
\bibitem{Cherednikov:2010uy}
  I.~O.~Cherednikov, A.~I.~Karanikas and N.~G.~Stefanis,
  Nucl.\ Phys.\  B {\bf 840} (2010) 379
  [arXiv:1004.3697 [hep-ph]].

\bibitem{Barone:2009hw}
  V.~Barone, S.~Melis and A.~Prokudin,
  Phys.\ Rev.\  D {\bf 81} (2010) 114026
  [arXiv:0912.5194 [hep-ph]].
  
   
  \bibitem{Burkardt:2004ur}
  M.~Burkardt,
  Phys.\ Rev.\  D {\bf 69} (2004) 091501;
  Phys.\ Rev.\  D {\bf 69} (2004) 057501

\bibitem{Pasquini:2010af}
  B.~Pasquini and F.~Yuan,
  Phys.\ Rev.\  D {\bf 81} (2010) 114013
  [arXiv:1001.5398 [hep-ph]].

  
  
\end{thebibliography}
\end{document}